\documentclass[12pt,draftcls,onecolumn] {IEEEtran}

\usepackage{stmaryrd}
\usepackage{mathrsfs}
\usepackage{amsmath}
\usepackage{amssymb}
\usepackage{multirow}
\usepackage{dsfont}
\usepackage{cite}
\usepackage[dvips]{graphicx}
\usepackage{amsmath}
\usepackage[ruled]{algorithm2e}
\usepackage{float}

\newcommand{\qed}{\nobreak \ifvmode \relax \else
      \ifdim\lastskip<1.5em \hskip-\lastskip
      \hskip1.5em plus0em minus0.5em \fi \nobreak
      \vrule height0.75em width0.5em depth0.25em\fi}

\begin{document}\pagestyle{empty}

\title{Antenna Deployment Method for MIMO Radar under the Situation of Multiple Interference Regions}
\author{Tianxian Zhang, Jiadong Liang, Yichuan Yang, Guolong Cui, Lingjiang Kong, and Xiaobo Yang
\thanks{Tianxian Zhang, Jiadong Liang, Yichuan Yang, Guolong Cui, Lingjiang Kong, and Xiaobo Yang are with the School of
Electronic Engineering, University of Electronic Science and Technology of
China, Chengdu, Sichuan, P.R. China, 611731. Fax: +86-028-61830064, Tel:
+86-028-61830768, E-mail: tianxian.zhang@gmail.com.
The research was supported by the National Natural Science Foundation of China under Grant
61501083 and the Fundamental Research Funds for the
Central Universities under Grant ZYGX2015KYQD056.}}

\maketitle

\begin{abstract}
In this paper, considering multiple interference regions simultaneously, an optimal antenna deployment problem for distributed Multi-Input Multi-Output (MIMO) radar is investigated. The optimal antenna deployment problem is solved by proposing an antenna deployment method based on Multi-Objective Particle Swarm Optimization (MOPSO). Firstly, we construct a multi-objective optimization problem for MIMO radar antenna deployment by choosing the interference power densities of different regions as objective functions. Then, to obtain the optimal deployment result without wasting time and computational resources, an iteration convergence criterion based on interval distance is proposed. The iteration convergence criterion can be used to stop the MOPSO optimization process efficiently when the optimal antenna deployment algorithm reaches the desired convergence level. Finally, numerical results are provided to verify the validity of the proposed algorithm.
\end{abstract}


\section{Introduction}\label{introduction}

Recently, distributed Multi-Input Multi-Output (MIMO) radar has attracted the attention of many researchers due to its advantages of spatial diversity and multiplexing gain for improving performances in many respects \cite{Haimovich,Fishler,He,Wang,Lehmann,Cao,Xia,Aggarwal,Godrich,Godrich2}. However, the performance advantages of such radar systems usually significantly rely on the positions of their antennas \cite{Godrich,Godrich2}. Thus, it is necessary to ascertain the optimal positions of antennas.

In these years, some researchers have focused on the problem of antenna deployment for MIMO radar \cite{placement1,placement2,Yang}. The problem is researched theoretically by considering only one single resolution cell \cite{placement1}. However, in practice, we usually need to interfere at least one large area, which includes many resolution cells. By discretizing the antenna deployment region into multiple small grids, a sequentially exhaustive enumeration (SEE) method is proposed to solve the antenna deployment problem \cite{placement2}. Nevertheless, the computation load of SEE will be exceedingly huge in real application. By using particle swarm optimization (PSO) in \cite{Yang}, it reduces the computational load and improves the optimized performance of a single surveillance region. But the above-mentioned papers only focus on the optimization performance of a single region. Moreover, they not only ignore the situation of multiple interference regions, but also neglect that additional time and computational resource may be wasted during the optimization process.

In real radar applications, situations and tasks always change when some emergency occurs. A MIMO radar system needs to react with current tasks quickly and ascertain the optimal position of antennas within a short duration. It means that multiple objective functions along with constraints always change fast and need to be solved in a short time. Meanwhile, time and computational resources are usually limited in real application. Therefore, it is important to obtain the optimal positions using time and computational resources as less as possible.

However, the most common way to stop the antenna deployment optimization algorithms (i.e., Multi-Objective PSO (MOPSO)) is simply to halt the execution when the algorithm has reached an artificial maximum iteration number \cite{Yang}. 
However, the maximum iteration number is difficult to determine. To be specific, \textbf{first}, if the maximum iteration number is set to be an over small number, the iterations will stop before the antenna deployment algorithm reaches the desired convergence level. \textbf{Second}, when the maximum iteration number is set to be an over large number, the iteration process will still continue even though no significant improvement can be obtained. In this case, additional time and computational resources will be wasted. Therefore, to obtain an optimal antenna deployment performance by using time and computational resources as less as possible, an iteration convergence criterion need to be investigated and the iteration should be stopped adaptively once the criterion is satisfied. Nevertheless, the iteration convergence criterion for optimal antenna deployment for MIMO radar remains largely unexplored, and it is the main topic of this paper.

In this paper, under the situation of multiple interference regions, an optimal antenna deployment problem for a distributed MIMO radar is studied. Firstly, considering multiple interference regions simultaneously, to evaluate interference performance of the MIMO radar, we choose the interference power density as objective functions. Then, in order to obtain optimal antenna deployment without wasting time and computational resources, a calculated method of interval distance and an iteration convergence criterion are proposed. Finally, numerical results are provided to verify the validity of the proposed iteration convergence criterion.

\section{Problem Formulation}\label{sect:multipath model}
Assume that a MIMO radar system contains $ J $ widely separated antennas. The antennas can be placed in some certain region, which is called as the deployment region. Assume that we need to interfere $ M $ regions simultaneously, these regions are defined as interference regions. The interference performances of the interference regions are important indexes to evaluate the performance of radar deployment. Generally, the interference power density is usually used as a criterion to evaluate the performance of interference.

Without loss of generality, we assume that the transmitted power of the $ j $th antenna is $ P_{tj} $ , and the Euclidean distance between the $ l $th resolution cell and the $ j $th antenna is $ R_{lj} $. Therefore, the interference power density of the $ l $th resolution cell in the $ m $th interference region is given by \cite{Richards,Skolnik,Richards10}
\begin{equation}
P_{l,m}(\Theta)=\sum_{j=1}^{J}\frac{P_{tj}G_{j}}{4\pi R_{lj}^2},
\end{equation}
where $ {\Theta}=\left[\theta_1^T,\theta_2^T,\dots,\theta_J^T\right] $ is a vector that contains the positions of all antennas, $(\cdot)^T$ denotes the transpose. $ G_j $ is the transmitting gain of the $ j $th antenna. $ \theta_j=\left[x_j,y_j\right]^T $ is the position of the $ j $th antenna. As we know, the interference performance of the $ m $th interference region is usually determined by the lowest interference power density within all resolution cells \cite{Richards,Skolnik,Richards10}. If we improve the lowest interference power density within all resolution cells in the $ m $th interference region, the interference performance of the $ m $th interference region will also be improved accordingly. Therefore, we choose the lowest interference power density within all resolution cells in the $ m $th interference region as an object. Then, the objective function $ f_m(\Theta) $ of the $ m $th interference region is defined as
\begin{equation}\label{equ:obj}
f_m({\Theta})=\min\limits_{0\leq l\leq{L_m}}\left\{P_{l,m}(\Theta)\right\},
\end{equation}
where $ L_m $ is the number of resolution cells within the $ m $th interference region.

Because $ M $ interference regions need to be interfered simultaneously, the antenna deployment problem is a multi-objectives optimization problem and the joint objective function can be described as
\begin{equation}
 {\mathcal{F}}_{M}(\Theta)=\left(f_1(\Theta),f_2(\Theta),\dots,f_M(\Theta)\right).	
\end{equation}

 Obviously,
 a bigger interference power density means a better interference performance for antenna deployment. Therefore, to obtain bigger interference power density $ {\mathcal{F}}_{M}(\Theta) $ of all the interference regions, we need to find a set of the best positions $ \hat{\Theta}$ of antennas, which can be described as
\begin{equation}
\hat{\Theta}=\arg \max \mathcal{F}_M(\Theta).
\end{equation}

This optimization problem of antenna deployment for MIMO radar is a high-dimensional and multi-objects problem. The MOPSO is an effective tool for this kind of optimization problem \cite{Aziz}. However, as is mentioned in Section \ref{introduction}, the situation and tasks always change quickly, if the iteration process of MOPSO still continues when no significant improvement can be obtained,
additional time and computational resource will be wasted. Therefore, the iteration should be stopped adaptively when the optimal antenna deployment algorithm reaches the desired convergence level. As a consequence, it is important for us to investigate the iteration convergence criterion for antenna deployment optimization algorithm, but this problem remains unexplored. The iteration convergence criterion will be proposed in the following section.

\section{MOPSO Algorithm for Radar Antenna Deployment and Iteration Convergence Criterion}\label{sect: conclusions}

This section contains two main parts: $ \boldsymbol{1)} $ for the challenge of high dimensionality, a deployment algorithm based on MOPSO for multiple interference regions is proposed. $ \boldsymbol{2)} $ To save time and  computational resources, an iteration convergence criterion based on interval distance for MOPSO is proposed.

\subsection{Antenna Deployment Algorithm based on MOPSO}

In multi-objective optimization problems, the objectives to be optimized are normally in conflict with respect to each other, which means an improvement in one of the objectives will probably result in a degradation of the other ones. Therefore, instead of finding a single optimal solution, we usually aim to find a group of non-dominated solutions that represent the best possible compromises among the objectives. In this paper, by using the MOPSO method, it can generate a group of non-dominated solutions, which constitute the Pareto front ($ \mathcal{PF} $) in a figure. Each non-dominated solution corresponds to an antenna deployment scheme. We can select a desired deployment scheme in preferential part of Pareto front according to different situations.

As $ M $ interference regions need to be interfered, there are $ M $ objective functions as in (\ref{equ:obj}) for $ m=1,2,\dots,M $. In the proposed optimization problem, each particle of MOPSO stands for a candidate deployment scheme. We suppose that the number of particles is $ N $. Then, the optimization process of MOPSO begins with a random initialization of these particles (i.e., for $n=1,2,\dots,N$, the position $ \Theta_n $ of the $ n $th particle are randomly set up within a fixed deployment region). After the initialization process, the joint objective function value $ \mathcal{F}_M(\Theta_n)=\left(f_1(\Theta_n),f_2(\Theta_n),\dots,f_M(\Theta_n\right) $ for each particle is evaluated.
As commonly setting in the MOPSO, an external archive is set up to store the particles whose corresponding values of objective function are non-dominated to each other. Thus, from the initial particles, we select the non-dominated ones out and store them in external archive. 

Then, all particles will be updated as follow in the $t$th iteration, for $  1\leq n \leq N $:
\begin{equation}
\begin{split}
v_n^{(t)}=&w{v_n}^{(t-1)}+c_1r_1\left(p_{n}^{(t-1)}-\Theta_n^{(t-1)}\right)\\
&+c_2r_2\left(p_{g}^{(t-1)}-\Theta_n^{(t-1)}\right),
\end{split}
\end{equation}
\begin{equation}
\Theta_n^{(t)}=\Theta_n^{(t-1)}+v_n^{(t)},
\end{equation}
where $ w $ is inertia weight, $ c_1 $ and $ c_2 $ are acceleration constants, $r_1$ and $ r_2 $ are random real values uniformly distributed in $ [0,1] $. $ v_n^{(t)} $ and $ \Theta_n^{(t)} $ respectively represent the current velocity and position of the $ n $th particle at the $t$th iteration. The velocity of particle in each dimension is set to be a random value within a fixed $V_{max}$. $ p_{n}^{(t)} $ is the individual best position of the $ n $th particle at the $t$th iteration (i.e, the best deployment scheme proposed by the $ n $th particle so far, and the initial position $ p_{n}^{(0)} $ is set to be equal as $\Theta_n^{(0)}$). $ p_{g}^{(t)} $ is the global best position at the $ t $th iteration (i.e, the best deployment scheme proposed by all particles so far), and is selected from the external archive with crowding distance selection methods \cite{Raquel,Coello2}.

After the update of position and velocity, we calculate the objective function of each particle, and update the individual best position $p_{n}^{(t)}$ of each particle as:
\begin{equation}
\!\!\!\!\!\!p_{n}^{(t)}=\begin{cases}
\Theta_n^{(t)},\ if\  \mathcal{F}_M\left(\Theta_n^{(t)}\right)\preceq\mathcal{F}_M\left(p_{n}^{(t-1)}\right)\\
p_{n}^{(t-1)} ,\ others
\end{cases},
\end{equation}
where $\mathbf{x}\preceq \mathbf{y}$ means that $\mathbf{x}$ is said to dominate\footnote{Taking a maximization problem as an example, given two vector $ \mathbf{x},\mathbf{y} \in \mathds{R}^Q$, we say that $ \mathbf{x} $ dominate $ \mathbf{y} $ (denoted by $ \mathbf{x}\preceq \mathbf{y} $) if and only if $ \mathbf{x} $ is partially greater than $ \mathbf{y} $, i.e., $ \forall q\in {1,\dots,Q}, {x}_q\geq {y}_q\wedge\exists q\in {1,\dots,Q}: {x}_q>{y}_q $ \cite{Coello2}.} $\mathbf{y}$.

 Then, we update the external archive, and the global best position (i.e, $ p_{g} $) will be selected out from the updated external archive again. Such iteration process will continue until the iteration convergence criterion is satisfied (this will be explained in the following section), where the optimization process will be stopped and the optimization results (i.e, the deployment schemes that stored in the external archive) will be outputted. Then, according to the practical situation, we can select a particle from the output for the MIMO radar antenna deployment.

\subsection{Iteration Convergence Criterion based on Interval Distance for MOPSO}
The most common way to stop the MOPSO iteration is simply to halt the execution when the algorithm has reached an artificial maximum iteration number \cite{Yang}. 
Nevertheless, the maximum iteration number is difficult to determine. Although the maximum iteration number can be obtained for simple problems in a quite straightforward way by trial and error, this procedure is computationally unaffordable for more complex practical problems.
Therefore, to obtain optimal performance by using time and resources as less as possible, the algorithm need to be stopped adaptively. Focusing on this issue, we proposed a novel iteration convergence criterion based on interval distance.

\begin{figure}[t]
	\centering
	\includegraphics[width=0.4\textwidth,draft=false]{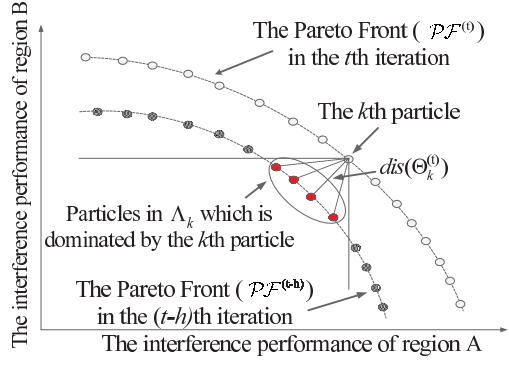}
	\caption{\!\!An example of the relative distance $dis(\Theta^{(t)}_k)$ with two objectives.}
	\label{fig:BP imaging diagrammatic sketch}
\end{figure}

In the iteration process of the MOPSO, it produces a $ \mathcal{PF} $ in each iteration \cite{Coello}. Meanwhile, the  $\mathcal{PF}$ will move outward with the improvement of the obtained non-dominated particles across different iterations as in Fig. \ref{fig:BP imaging diagrammatic sketch}. Therefore, we can decide whether the algorithm reaches desired convergence level by observing the movement of the $ \mathcal{PF} $ over iterations. In this paper, the interval distance is proposed to measure the distance between two Pareto fronts $\mathcal{PF}^{(t-h)}$  and  $\mathcal{PF}^{(t)}$ at the $(t-h)$th and $t$th iterations, respectively, as in Fig. \ref{fig:BP imaging diagrammatic sketch}. The proposed interval distance can be calculated for every calculation step $ h $, whose value can be set according to the practical situations. Obviously, the interval distance can be denoted as an index to describe the speed of interference performance improvement efficiently. The iteration will be stopped adaptively once the interval distance reaches desired value.

For the Pareto front $ \mathcal{PF}^{(t)} $ in the $t$th iteration, the objective function value of the $ k $th particle in $ \mathcal{PF}^{(t)} $ is denoted as $ {\mathcal{F}}_{M}(\Theta^{(t)}_k),\ 1\leq k\leq K^{(t)} $, where $ K^{(t)}  $ is the number of particles (or the number of non-dominated solutions) within $\mathcal{PF}^{(t)}$. The detail calculation method of the interval distance is described as following.

\textbf{Firstly}, for each particle within the Pareto front $\mathcal{PF}^{(t)}$, we sort out the particles within the Pareto front $\mathcal{PF}^{(t-h)}$ which is dominated by the $ k $th particle within the Pareto front $\mathcal{PF}^{(t)}$ and store them in a dominated external archive $ \Lambda_{k} $ as in Fig. \ref{fig:BP imaging diagrammatic sketch}. The value of the $ i $th particle in $ \Lambda_{k} $ is denoted as ${\mathcal{{F}}}_{M}(\Theta^{(t-h)}_i)$. Thus, we obtain
\begin{equation}\label{Firstly}
{\mathcal{F}}_{M}\left(\Theta^{(t)}_k\right)\preceq{\mathcal{{F}}}_{M}\left(\Theta^{(t-h)}_i\right),\ {\Theta^{(t-h)}_i}\in{\Lambda_{k}},\ 1\leq{i}\leq{|\Lambda_{k}|}.
\end{equation}

\textbf{Secondly}, we calculate the Euclidean distances from the $k$th particle within the Pareto front $\mathcal{PF}^{(t)}$ to all the particles within $\Lambda_k$ as in Fig. \ref{fig:BP imaging diagrammatic sketch}, and choose the minimum distance as the relative distance of the $k$th particle in $\mathcal{PF}^{(t)}$ to $\mathcal{PF}^{(t-h)}$, i.e., 
\begin{equation}
\label{realtive}
dis\left(\Theta^{(t)}_k\right)=\min\limits_{1\leq{i}\leq{|\Lambda_{k}|}}\left[\sum_{m=1}^M \left(f_m\left(\Theta^{(t)}_k\right)-f_m\left(\Theta^{(t-h)}_i\right)\right)^2\right]^{{1}/{2}}.
\end{equation}
If there is no particle in $ \Lambda_k $, $ dis(\Theta^{(t)}_k) $ will be set as $ 0 $.

\textbf{Then}, after calculating the relative distances $dis(\Theta^{(t)}_k)$, $1\leq k\leq K^{(t)}$, of all the particles within $\mathcal{PF}^{(t)}$, we can evaluate the outward movement of the Pareto front by further processing of the relative distance data. However, sometimes there are outliers, i.e., $dis(\Theta^{(t)}_k)=0$, which are need to be eliminated. We consider three processing method, such as maximization, minimization and average, i.e.,
\begin{equation}\label{maxdis}
\!\!\!\!\textbf{Maximization}: dist^{(t)}
=\!\!\!\!\max\limits_{1\leq{k}\leq K^{(t)},dis(\Theta^{(t)}_k)\neq{0}}{dis\left(\Theta^{(t)}_k\right)},
\end{equation}
\begin{equation}\label{mindis}
\!\!\!\!\textbf{Minimization}:dist^{(t)}
=\!\!\!\!\min\limits_{1\leq{k}\leq K^{(t)},dis(\Theta^{(t)}_k)\neq{0}}{dis\left(\Theta^{(t)}_k\right)},
\end{equation}
\begin{equation}\label{avgdis}
\textbf{Average}:dist^{(t)}
=\dfrac{\sum\limits_{k=1,dis(\Theta^{(t)}_k)\neq{0}}^{K^{(t)}}{dis\left(\Theta^{(t)}_k\right)}}{K^{(t)}-z},
\end{equation}
where $ z $ is the number of particles whose relative distance is zero.

\textbf{Finally}, we set an iteration convergence threshold. The iteration convergence criterion is described as below:
\begin{equation}\label{Delta}
|dist^{(t)}-dist^{(t-h)}|\leq\Delta\ ,
\end{equation}
where $ \Delta $ is a desired convergence threshold and can be set according to the practical situations. Notice that when the difference of interval distance between the current $t$th iteration and previous $(t-h)$th iteration is below a threshold $ \Delta $, it means that the antenna deployment performance has reached the desired convergence level. Therefore, the iteration can  be stopped and the optimization results will be output.
Summary of the proposed convergence criterion is given in Algorithm \ref{alg:method}.

\begin{algorithm}[htb]
	\DontPrintSemicolon
	\caption{\label{alg:method} Summary of iteration convergence criterion}
The MOPSO algorithm is proceed and iteration number $ t\geq h $.\;	
		\If{$ t=h,h+1,h+2,\dots$}{
			\For{the arbitrary $k$th particle in $\mathcal{PF}^{(t)}$}{
				Sort out the particles in $\mathcal{PF}^{(t-h)}$ which is dominated by the $k$th particle in $\mathcal{PF}^{(t)}$ and store them in $ \Lambda_{k} $. \;
				\For{the arbitrary $i$th particle in $ \Lambda_{k} $ as in (\ref{Firstly})}{
					Calculate the Euclidean distance between the $i$th particle in $ \Lambda_{k} $ and the $k$th particle in $\mathcal{PF}^{(t)}$.\;	
				}
				Choose the minimum Euclidean distance as the relative distance $ dis(\Theta^{(t)}_k) $.
			}
			Calculate the interval distance $ dist^{(t)} $ using (\ref{maxdis}) or (\ref{mindis}) or (\ref{avgdis}).\;
			\If {$ |dist^{(t)}-dist^{(t-h)}|\leq\Delta $}{
				Break the MOPSO algorithm.\;}
		}
Obtain the desired solutions.
\end{algorithm}

\section{Nmerical Results}
In this section, we demonstrate the advantages and validity of the proposed antenna deployment algorithm. Firstly,  we assume that two interference regions (region A and B, i.e., $M=2$) are to be interfered simultaneously. The radar transmitted power and transmitting gain are $ P_{tj}=15$ kW and $G_j=40$ dB, respectively, for $ j=1,\dots,J $. The simulation region is set as a square with size of $ 70 $ km $ \times $ $ 70 $ km. The number of antennas is $ J=8 $. According to (1) and (2), the objective function is obtained as $ \mathcal{F}_2(\Theta)=\left(f_1(\Theta),f_2(\Theta)\right) $, where $ f_1(\Theta) $ and $ f_2(\Theta) $ are the objective functions in region A and B, respectively. For MOPSO, the number of particles $ N=30$, $V_{max}=4$, $c_1=c_2=2$, $w=0.4$, $h=5 $. We also set the threshold $ \Delta $ in (\ref{Delta}) as $ 0.25\times10^{-3} $. To obtain the statistical performance, the results are obtained from
$1000$ independent Monte Carlo trials.

\begin{figure}[t]
	\begin{minipage}{0.5\linewidth}
		\centerline{\includegraphics[width=1\textwidth,draft=false]{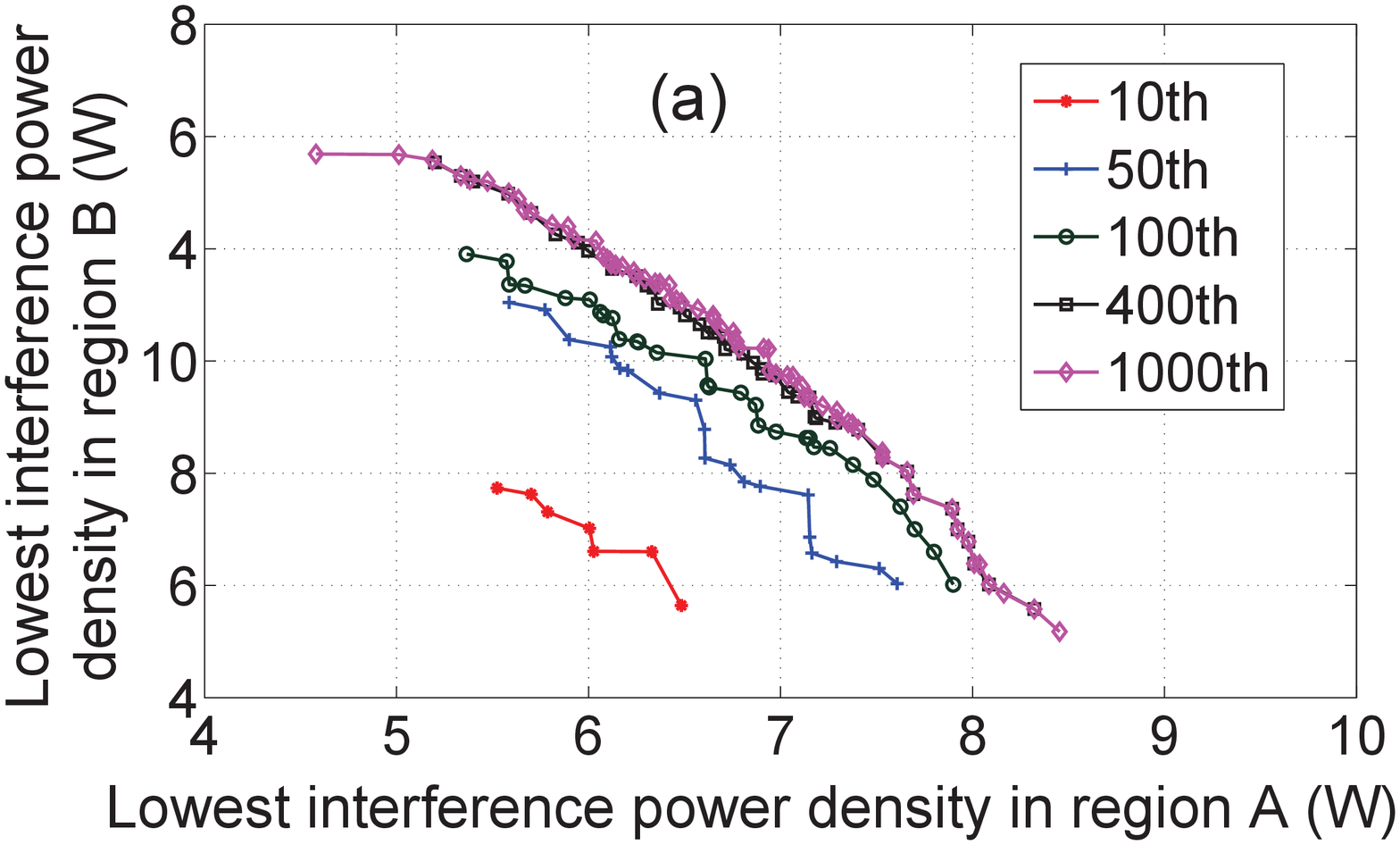}}
	\end{minipage}
	\hfill
	\begin{minipage}{0.44\linewidth}
		\centerline{\includegraphics[width=1\textwidth,draft=false]{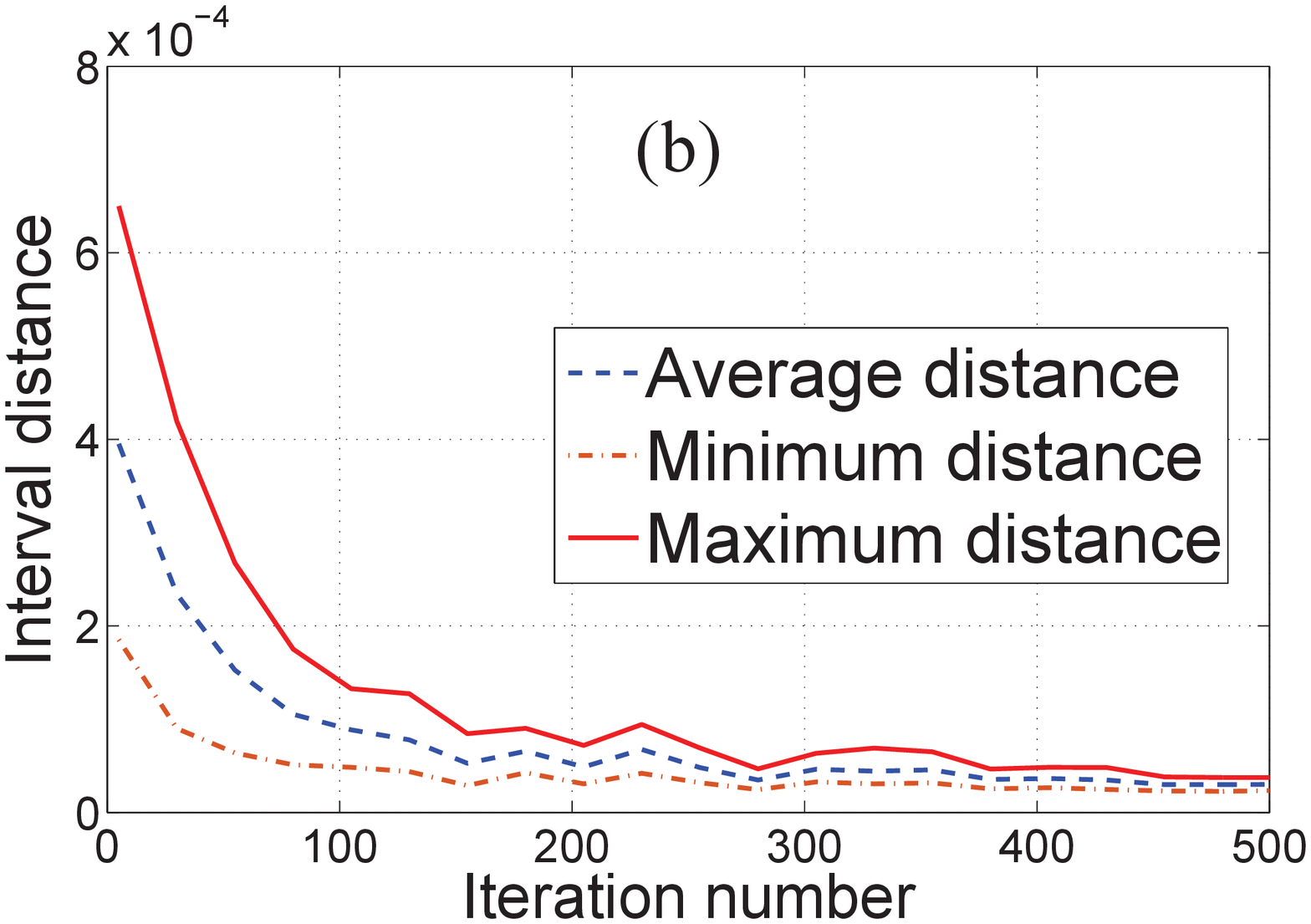}}
	\end{minipage}
	\vfill
	\caption{(a). Pareto front of the $ F_2(\Theta) $ in region A and region B; (b). Simulation results of interval distance over different iterations.}
	\label{fig: cac num2}
\end{figure}

We can see from Fig.\ref{fig: cac num2} (a) that with the increase of iteration number, the $ \mathcal{PF} $ moves outward fast at first and then moves more and more slowly. Finally, the curve seem to stop moving when the iteration number $t>400$. The curve of the Pareto front $ \mathcal{PF}^{(t)} $ for $t=400 $ is almost coincided with the curve of the $ \mathcal{PF}^{(t)} $ for $t=1000 $ iterations. Obviously, if the iteration process still continues after $t=400$, additional time and computational resources would be wasted. In Fig.\ref{fig: cac num2} (b), it is shown that all of the $3$ curves decline rapidly at the beginning, and then the curves do not decline obviously over iterations.
Meanwhile, when the iteration number is increased to $400$, the antenna deployment performance is tended to reach a desired convergence level and the MOPSO iteration could be stopped adaptively. It is consistent with the results in Fig.\ref{fig: cac num2} (a).

	\begin{table}
		\newcommand{\tabincell}[2]{\begin{tabular}{@{}#1@{}}#2\end{tabular}}
		\centering
		\caption{}
		\label{table1}
		\begin{tabular}{|c|c|c|c|c|c|c|}
			\hline
			\multirow {2}{*}{\tabincell{c}{Iteration\\ Number}}
			&\multicolumn{2}{|c|}{$ f_1(\Theta)=5.52\textrm{W} $}
			&\multicolumn{2}{|c|}{$ f_1(\Theta)=6.03\textrm{W} $}
			&\multicolumn{2}{|c|}{$ f_1(\Theta)=6.48\textrm{W} $}\\\cline{2-7}
			&$ f_2(\Theta) $&$ c(\%) $&$ f_2(\Theta) $&$ c(\%) $&$ f_2(\Theta) $&$ c(\%) $\\
			\hline
			$ 10th $&5.86W &68.9 &5.30W &67.4 &4.82W &65.4 \\
			\hline
			$ 50th $&7.52W &88.5 &7.04W &89.6 &6.13W &83.1\\
			\hline
			$ 100th $&7.68W &90.3 &7.38W &93.9 &6.77W &91.7\\
			\hline
			$ 400th $&8.49W &99.8 &7.82W &99.5 &7.25W &98.3\\
			\hline
			$ 1000th $&8.50W &100 &7.86W &100 &7.37W &100\\
			\hline
		\end{tabular}
	\end{table}
	
	\begin{table}
		\newcommand{\tabincell}[2]{\begin{tabular}{@{}#1@{}}#2\end{tabular}}
		\caption{}
		\label{table2}
		\centering
		\begin{tabular}{|c|c|c|c|c|c|c|}
			\hline
			\multirow {2}{*}{\tabincell{c}{Iteration\\ Number}}
			&\multicolumn{2}{|c|}{$ f_2(\Theta)=4.82\textrm{W} $}
			&\multicolumn{2}{|c|}{$ f_2(\Theta)=5.30\textrm{W} $}
			&\multicolumn{2}{|c|}{$ f_2(\Theta)=5.86\textrm{W} $}\\\cline{2-7}
			&$ f_1(\Theta) $&$ c(\%) $&$ f_1(\Theta) $&$ c(\%) $&$ f_1(\Theta) $&$ c(\%) $\\
			\hline
			$ 10th $&6.48W &80.2 &6.03W &77.5 &5.52W &69.9 \\
			\hline
			$ 50th $&7.60W &94.5 &7.16W &89.6 &7.14W &90.5\\
			\hline
			$ 100th $&7.90W &97.7 &7.78W &92.0 &7.62W &96.5\\
			\hline
			$ 400th $&8.05W &99.6 &7.74W &99.2 &7.84W &99.4\\
			\hline
			$ 1000th $&8.08W &100 &7.80W &100 &7.89W &100\\
			\hline
		\end{tabular}
	\end{table}	

To further compare with the interference performance (i.e., power density) over different iterations, we randomly fix the performances of one interference region, and observe the performance changes of another interference region over iterations. We define $ c $ as the ratio of the performance of current iteration to the performance for the iteration $t=1000$. The interference performances for region A and region B are shown in TABLE \ref{table1} and TABLE \ref{table2}, respectively. The results indicate that, when the iterations reach $400$, the interference performances are almost the same as the performances for the iteration $t=1000$. These phenomenons demonstrate the validity of the proposed iteration convergence criterion and antenna deployment algorithm.

\section{Conclusion}
In this paper, considering multiple interference regions simultaneously, an optimal antenna deployment problem for distributed MIMO radar has been studied. The optimization goal is to obtain the optimal interference performance by using MOPSO method without wasting time and computational resources. We have proposed an iteration convergence criterion based on interval distance. The interval distance could be used to present the improvement of different iterations effectively. The iteration would be stopped adaptively once the criterion is satisfied. Finally, the simulation results show that the proposed algorithm can achieve the optimal deployment results effectively.

\end{document}